\documentstyle[prd,aps]{revtex}

\begin{document}
\tighten
\draft
\twocolumn[\hsize\textwidth\columnwidth\hsize\csname 
@twocolumnfalse\endcsname
\title{Reply to `Comment on ``The influence of cosmological transitions 
on the evolution of density perturbations'' '}
\author{J\'er\^ome Martin}
\address{DARC, Observatoire de Paris, \\ 
UPR 176 CNRS, 92195 Meudon Cedex, France. \\
e-mail: martin@edelweiss.obspm.fr}
\author{Dominik J. Schwarz}
\address{Institut f\"ur Theoretische Physik, \\
Robert-Mayer-Stra\ss e 8 -- 10, Postfach 11 19 32, \\ 
D-60054 Frankfurt am Main, Germany. \\
e-mail: dschwarz@th.physik.uni-frankfurt.de}
\maketitle

\begin{abstract}
We reply to L.P. Grishchuk's comment \cite{GCom} on our paper ``The 
Influence of Cosmological Transitions on the Evolution of Density 
Perturbations'' \cite{MS}. We show that his points of criticism are not 
correct. 
\end{abstract}
\narrowtext
\vspace{1 cm}]

Inflationary cosmology predicts the generation of density perturbations 
and gravitational waves with almost scale-invariant spectra. The ratio 
between their amplitudes (at superhorizon scales) is fixed by the 
equation of state during inflation. The closer the inflationary epoch 
is to a de Sitter phase, the larger are the generated density perturbations 
with respect to the gravitational waves. This statement is the so-called 
``standard result''. The reason for this result is that
the gauge-invariant metric perturbation $\Phi$ grows by a huge factor 
during reheating, whereas the gravitational waves amplitude $h_{\rm gw}$ 
essentially remains constant during that epoch. 

Four years ago Grishchuk published a paper \cite{G} in which he stated 
that there is no big amplification of density perturbations. As 
a consequence, he expressed his disagreement 
with the standard result. This claim was criticized by Deruelle and 
Mukhanov \cite{DM}. 
They wrote that Grishchuk had not properly taken the joining conditions 
at reheating into account. A central point of our paper \cite{MS} is 
that Grishchuk's criticism was wrong indeed (and that the standard result 
is correct), not because of the joining conditions, though. The origin 
of Grishchuk's error was that he had not evolved the perturbations through 
reheating correctly.
\par
Let us draw our attention on four issues that are discussed in 
Grishchuk's comment: 
\par
{\bf 1. Sharp transition:}  
Grishchuk defines a new quantity $\bar{\mu }:=\mu /\sqrt{\gamma }$. 
The introduction of this new variable $\bar{\mu }$ does not add
anything new to the problem. In particular, it does not alter 
the behavior of the growing superhorizon mode 
$\mu \propto a\sqrt{\gamma }$, see the equation before Eq.~(14) in 
Ref.~\cite{GCom}.
\par
Below Eq.~(17) of Ref.~\cite{GCom} Grishchuk states that $\bar{\mu}$ 
is continuous through a sharp transition of the equation of state.
This is true and we have used this condition in Eq.~(6.5) of our paper. 
In addition, Grishchuk writes that ``It was shown \cite{G,G2b} that 
Eq.~(9) requires the continuity of the function $v$, where 
$v=\gamma (\bar{\mu }/a)'$ and, hence, the continuity of the function 
$\bar{\mu }/a$.'' and ``This evolution is at the basis of practical 
calculations in \cite{G,G2b,G3}''. This is in fact in total contradiction 
to the ``practical calculations'' done in the quoted Ref.~\cite{G}. 
Firstly, the quantity $\bar{\mu }$ is not even defined in that paper and 
therefore cannot have been used. Secondly, one of the equations after 
Eq.~(48) (on p.~7161) is:
\begin{equation}
\label{1}
\mu |_{\eta =\eta _1-0}=\mu |_{\eta =\eta _1+0}.
\end{equation}
Therefore, it is clear that Grishchuk used the continuity of $\mu $ and 
not that of $\bar{\mu }$. Now Grishchuk says that this equation is a 
misprint \cite{Private}, which we do not believe. On the contrary, we 
think that this is the origin of the error as it is explained in our paper.
If one uses the continuity of $\mu $ (mistakenly), then one loses the factor 
$\sqrt{\gamma }$, which is responsible for the large amplification. This 
was exactly the mistake in Ref.~\cite{G}. 
\par
{\bf 2. Constancy of $\zeta$:}
The arguments in \cite{GCom} concerning the constancy of $\zeta $ are 
inconsistent. Grishchuk writes that 
``The mysterious `$\zeta $ conservation law' 
has only that meaning that the `growing' part of the function 
$\bar{\mu }/a$, in its lowest nonvanishing long-wavelength 
approximation, is a constant ...''. This is confirmed by 
Eq.~(4.21) of our paper and is one of the conclusions of our article. 
If the decaying mode is neglected, one concludes that 
$\zeta =-\bar{A}_1/2=cte$. $\zeta$ is a constant and is determined from 
the initial conditions. But ten lines later, Grishchuk writes the contrary: 
``It was shown \cite{G2b,G3} that if you dare to respect the original 
Einstein equations, the constant (23), must be a strict zero, and not a 
constant determined by initial conditions''!
\par
Obviously, one is free to use other methods than the $\zeta$ argument
to calculate the evolution of the perturbations. But if one chooses to 
use the constancy of $\zeta $, the result will be correct (see Ref.~\cite{MS}).
\par
{\bf 3. Big amplification:} 
Grishchuk's words do not say the same as his equations. He says that 
``big amplification is a misinterpretation'' . But, ironically, he 
himself derives
the standard result (as it is obvious if one calculates the ratio of the 
Eqs.~(33) and (36) of Ref.~\cite{GCom})! We do not understand how Grishchuk 
can say, at the same time, that the standard result is wrong and that he 
agrees with Eq.~(4.25) of our work \cite{MS} (bottom of p.~10), since this 
equation is precisely the standard result! 

Grishchuk gives an example to show that if the amplification coefficient 
is big, it does not follow that the real (absolute) value of 
$\Phi (\eta _m)$ is big also. Again this remark is out of place. We wrote 
exactly the same after Eq.~(4.7) in our article. In order to determine
the absolute value one needs to specify the initial conditions.
\par
{\bf 4. Quantum initial conditions:}
Finally, the quantum normalization of density perturbations used on p.~12 of 
Ref.~\cite{GCom} is wrong. Apparently, in his comment Grishchuk changes 
his opinion and now accepts the discontinuity in $\mu$ but, in order 
to avoid the factor $\sqrt{\gamma }$ again, uses wrong initial conditions. 
Grishchuk claims that the scalar metric perturbation $h$ in the synchronous 
gauge is $h(k) \approx (l_{\rm Pl}/l_0) k^{2 + \beta}$. The last 
equation implies the asymptotic behavior $\lim _{k \rightarrow +\infty }
\bar{\mu}(\eta,{\bf k})=-4\sqrt{\pi}l_{\rm Pl} e^{-ik(\eta -\eta _0)}/
\sqrt{2k}$. Together with the (correct) normalization of the scalar field 
$\lim _{k \rightarrow +\infty } \varphi_1(\eta,{\bf k})=
\sqrt{\hbar}e^{-ik(\eta -\eta _0)}/ (a\sqrt{2k})$ (see Sec.~VII of 
Ref.~\cite{G}), the normalization of $\bar\mu$ is in contradiction 
to the Einstein equations. 
The normalization of the scalar field fixes the normalization of density 
perturbations via the $0-i$ component of the Einstein equations:
\begin{equation}
\label{2}
\lim _{k \rightarrow +\infty }\mu (\eta ,{\bf k})=-\sqrt{2\kappa }a
\varphi _1(\eta ,{\bf k}) \ .
\end{equation} 
Since $\mu $ appears in this equation, and not $\bar{\mu}$, this leads 
to the following correct asymptotic behavior (as it is demonstrated in the 
appendix of our paper \cite{MS}): $\lim _{k \rightarrow +\infty }\mu 
(\eta ,{\bf k})=-4\sqrt{\pi }l_{Pl}
e^{-ik(\eta -\eta _0)}/\sqrt{2k}$, which implies 
$h(k) \approx (l_{\rm Pl}/l_0) k^{2 + \beta}/\sqrt{\gamma}$. Again, 
Grishchuk is wrong by a factor $\sqrt{\gamma }$. 
\par
In conclusion, Grishchuk's points of criticism \cite{GCom} on 
our paper \cite{MS} are incorrect. His remarks are either inconsistent 
or an (involuntary) confirmation of the results obtained in 
Ref.~\cite{MS}. We consider this controversy to be settled now.

\section*{Acknowledgements}

D.J.S. thanks the Alexander von Humboldt foundation for a fellowship.

\end{document}